# Coherent Slowing of a Supersonic Beam with an Atomic Paddle

E. Narevicius,[1] A. Libson,[1] M.F. Riedel,[1] C.G. Parthey,[1] I. Chavez,[1] U. Even,[2] and M.G. Raizen[1,*]

[1]*Center for Nonlinear Dynamics and Department of Physics, The University of Texas at Austin, Austin, Texas 78712-1081, USA*

[2]*Sackler School of Chemistry, Tel-Aviv University, Tel-Aviv, Israel*

(Dated: December 8, 2006)

## Abstract

We report the slowing of a supersonic beam by elastic reflection from a receding atomic mirror. We use a pulsed supersonic nozzle to generate a $511 \pm 9$ m/s beam of helium that we slow by reflection from a Si(111)-H(1x1) crystal placed on the tip of a spinning rotor. We were able to reduce the velocity of helium by 246 m/s and show that the temperature of the slowed beam is lower than 250 mK in the co-moving frame.

[*]Electronic address: `raizen@physics.utexas.edu`

Atomic and molecular beams have been a fundamental tool of physical chemistry for many years, with wide applications in the study of gas dynamics and surface interactions [1, 2]. In recent years, they have also been used as the source for atom optics experiments with potential applications in atom lithography and atomic interferometry as well as fundamental physics and precision tests [3]. The highest brightness atomic or molecular beam is formed when gas escapes from a high pressure source through an aperture that is large compared to the mean-free-path in the gas. The remarkable result known as a supersonic beam combines intensities of $10^{23}$ atoms/sr/s with beam temperatures of 50 mK in the moving beam frame. The main limitation of the supersonic beams is that the mean velocity in the laboratory frame is quite high which has meant decreased sensitivity and energy range for surface scattering experiments, and greater difficulties controlling the beam in atom optics experiments. Experimental realizations of slow supersonic beams include work done by Gupta and Herschbach [4]. They controlled the beam velocity by mounting a continuous flow supersonic valve on the tip of a rotor. Bethlem et al. showed slowing of polar molecules using pulsed electric fields [5]. Optical pulsed fields have been used for slowing as was demonstrated by Fulton et al. [6]. A method to stop molecules was also demonstrated by Elioff et. al. where two crossed beams produced cold NO molecules by billiard like collisions [7].

Motivated by the success of the neutron beam paddle [8] we have built an atomic paddle that reduces the mean velocity of the supersonic beam by specularly reflecting the atoms from an atomic mirror moving in the beam direction. Supersonic beam velocities typically range from 250 m/s to several kilometers/s depending on the temperature and make-up of the gas. In the idealized case of normal reflection from a linearly moving crystal, the reflected velocity will be changed by twice the velocity of the mirror. Specifically, we see that $v_f = -v_a + 2v_c$, where $v_a$ is the initial velocity of the gas, $v_f$ is the final velocity, and $v_c$ is the crystal velocity. From this, it is clear that mirror velocities in excess of 100 m/s are desirable, which can be achieved using rotary motion.

It is well known that helium reflects from single crystal surfaces and helium diffraction was first used by Stern and Estermann in 1930 to measure the Maxwellian velocity distribution [9]. Recently, Allison et. al. were even able to focus a supersonic beam of helium using reflection from a curved hydrogen-passivated Si (111) wafer [10, 11]. To elaborate on the important point of helium reflection probability we consider helium atoms impinging on the surface of a stationary single-crystal material. The fraction of atoms that undergo elastic

reflection is given by

$$I/I_0 = e^{-2W} \tag{1}$$

where $W$ is the Debye-Waller factor

$$2W = \frac{24mE_{0z}T_S}{Mk_B\theta_S^2}. \tag{2}$$

Here $m$ is the mass of the impinging atom, $M$ is the mass of the atom in the crystal, $T_S$ is the temperature of the crystal, $E_{0z}$ is the incident energy, and $\theta_S$ is the crystal Debye temperature [12, 13]. For example, consider $^4$He impinging on a single-crystal $^{28}$Si with incident velocity of 550 m/s. The Debye temperature in this case is 690 K, and even for room temperature crystals we find reflection probabilities above 25%. Allison et. al. have measured a specular reflection probability of 3% for a Helium beam at a normal velocity of 900 m/s [14]. This number is consistent with the Debye-Waller prediction for elastic reflection, as the flux was distributed among over 40 diffraction channels. Note that in our experiment, the impact velocity in the frame of the moving crystal is lower than the initial velocity of the supersonic beam, which helps to increase the elastic reflection probability even more.

A schematic figure of our experimental setup is presented in Fig. 1. Our pulsed supersonic valve generates an atomic beam that is collimated by a single skimmer. The collimated beam reflects from a spinning rotor and the slow beam is detected by a quadrupole mass spectrometer (QMS).

Our atomic beam source has been developed by Even and Lavie [15, 16] and further optimized to fit our experiment [17]. The trumpet shaped nozzle creates a supersonic beam of pure helium that is very directional (half-angle of seven degrees) and very monochromatic (less than 1 percent relative spread in velocity). It combines pulse durations as short as $10\,\mu$s FWHM (repetition rate up to 40 Hz) with cryogenic operation. The valve can operate at up to 100 atmospheres backing pressure. As a result of these features, we are able to time the valve so that we only "fire" when the crystals are aligned with the incoming beam and so no atomic flux is wasted between the pulses, which would raise the background pressure and possibly attenuate the beam. Two compact 300 l/s turbomolecular pumps maintain a pressure of $10^{-7}$ Torr in the source chamber.

Collimated by a 5 mm skimmer, the supersonic beam then enters the rotor chamber, which is kept at a pressure of $10^{-8}$ Torr by a 500 l/s turbomolecular pump. The chamber contains a 50.4 cm radius titanium rotor coupled to a motor via a ferrofluidic feedthrough that can operate at up to 10,000 RPM. The rotation rate is servo-controlled to within a few thousandths of one Hertz of the desired rotation rate. We synchronize the supersonic beam arrival time with the rotor position using a pulse generated by the rotary motion encoder that has an angular resolution better than 1 mrad.

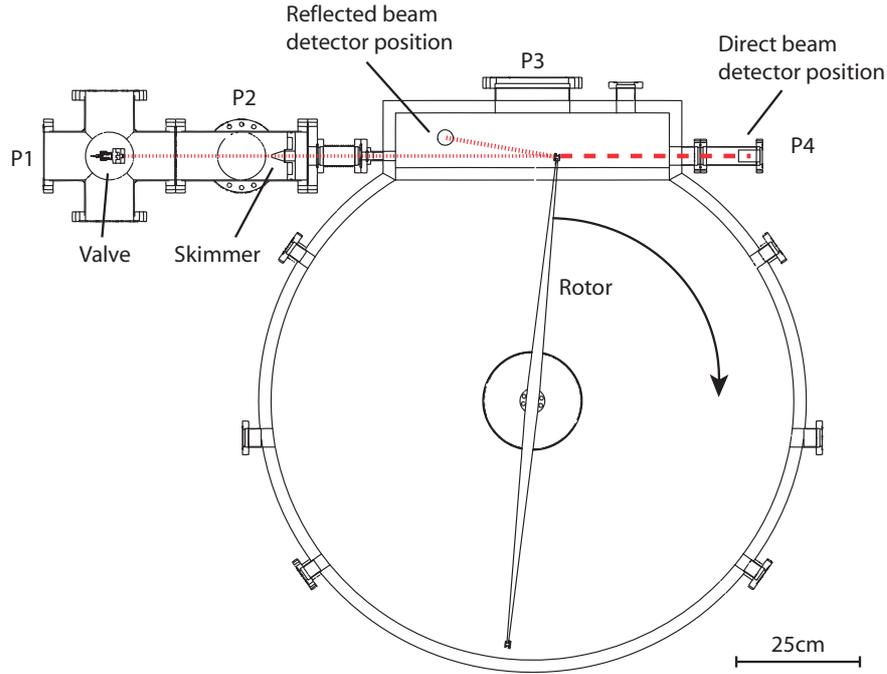

FIG. 1: Schematic of our apparatus . Positions of turbomolecular pumps are indicated as P1-P4; the slowed atom trajectory is shown as a dotted line. The dashed line stands for the propagation path of the direct beam.

Rotary motion, though, creates an unwanted fanning effect. Due to a finite nozzle opening time, the temporal extent of our atomic beam arriving at the spinning mirror is about $115\mu$s. Over this period of time the angle of the reflecting mirror changes in proportion to the angular velocity, and the reflected atomic beam is spread in the rotation plane. We use a large diameter rotor because it allows lower rotation rates, helping to reduce the fanning effect. However, the tensile stress induced by centrifugal forces sets a safety limit on the rotor length. In our case the tensile stress is 4 times lower than the yield tensile strength for the titanium alloy used, as was confirmed by finite element analysis.

Two 9 mm diameter Si(111)-H(1x1) atom mirrors are mounted in the crystal holders at the ends of the rotor. The crystals are prepared ex-situ by wet etching a silicon wafer in 40% ammonium fluoride. This process removes the native oxide layer and the dangling silicon bonds are passivated with hydrogen [18]. We routinely verify the surface preparation quality using an atomic force microscope.

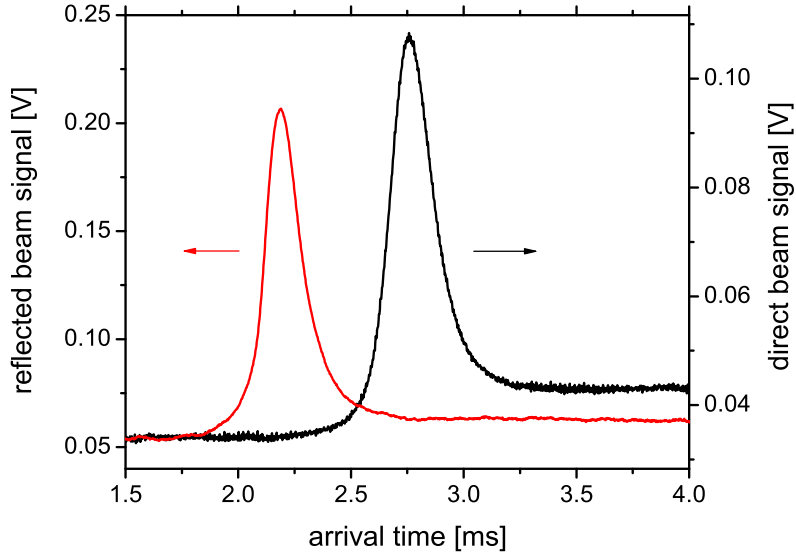

FIG. 2: This plot shows the direct helium beam (right peak) and the beam reflected from the still rotor (left peak). The reflected beam arrives sooner because the distance the beam travels is less than for the direct beam. Our calculations show that both beams have the same speed and temperature. We cannot compare the amplitudes of these beams, as the data was taken using different instruments, and the detectors were mounted in different configurations.

We detect the helium beam using two QMSs equipped with an electron multiplier output [19]. The short pulse duration allows us to make time-of-flight measurements for both the direct and the reflected beams.

In Fig. 2 we present the time-of-flight measurement for the direct and reflected atomic beams. We use a 48:52 mixture of helium and neon cooled to 77 K. Seeding helium with a heavier atom helps us to reduce the initial velocity of the incoming supersonic beam. As one can see, the initial 10 $\mu$s long gas pulse spreads to 200 $\mu$s after traveling 1.42 m to the detector. The pulse dispersion is the direct consequence of the finite beam temperature. We extract the arrival time and the distribution width from the time-of-flight data and evaluate the mean velocity of the beam, $v_a = 511 \pm 9$ m/s, and the most probable velocity in the moving frame, $v_{mp} = 18.4 \pm 0.5$ m/s. The temperature of the beam can be calculated from the beam speed ratio which is the ratio of the mean to the most probable velocities [2]. In our case the speed ratio, $s = \frac{v_a}{v_{mp}} = 27.7 \pm 1.9$ corresponds to a temperature of $249 \pm 36$ mK.

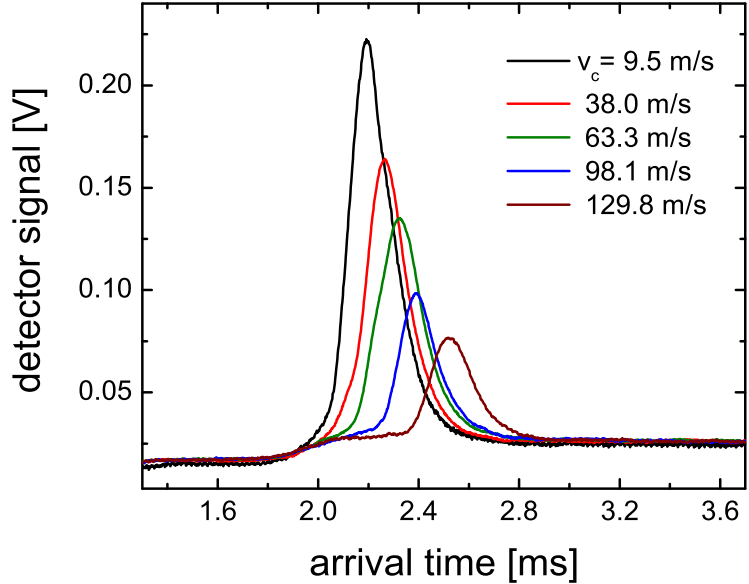

FIG. 3: The QMS signal of the helium beam reflected from the spinning rotor for different crystal velocities ($v_c$). The later arrival time corresponds to a slower beam. The lower amplitude of the signal for higher rotor velocities is due to the fanning effect, and agrees with our simulations. These curves are averages over 50 data sets, and this artificially broadens the peaks as they are averages over sharp peaks with varying arrival time due to a jitter in rotor trigger position.

In order to measure the reflected beam signal, we align the rotor 5° degrees off the perpendicular position with respect to the incoming atomic beam (as shown in Fig. 1). As one can see in Fig. 2 the arrival time and distribution width become smaller due to the shorter 1.11 m travelling distance from the valve to the reflected beam detector. Importantly, the temperature of the reflected beam is $254 \pm 41$ mK. This result confirms our expectation that the specular reflection does not significantly change the velocity distribution of the reflected beam.

We present time-of-flight data of the beam reflected from the moving mirror in Fig. 3. As one can see, the reflected beam arrival time is delayed with increasing mirror linear velocity.

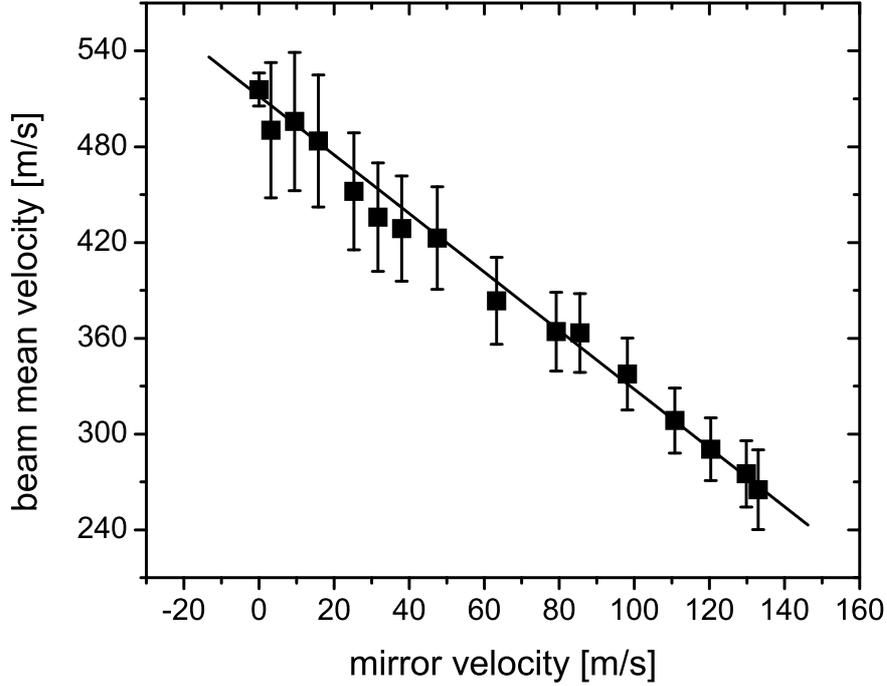

FIG. 4: The calculated velocity of the slowed beam, plotted against the velocity of the crystal.

The rotor fanning effect reduces the intensity of the reflected slow beam in quantitative agreement with our simulations. The beam slowing results are summarized in Fig. 4 where we present the mean velocity of the reflected beam as a function of the crystal linear velocity. We are able to subtract up to 246 m/s from the initial velocity of the incoming beam. Note that we can also accelerate the supersonic beam by reversing the rotation direction. Currently, we are able to remove 246 m/s from the initial beam velocity since we are limited to a maximal rotation frequency of 42 Hz. Further improvements in rotor design (increased shaft torsional stiffness) should enable us to at least double this number.

Increasing the rotor diameter is not the only way to compensate for the flux loss due to the fanning effect. In our Monte Carlo beam simulations [17] we have investigated the possibility of using a deformed crystal surface as a concave atomic focusing mirror. We have seen that the focusing increases our predicted flux values by an order of magnitude. For a pure helium beam we expect fluxes of $10^{13}$ atoms/s at velocity of 300 m/s. We believe that we will be able to reach this number after redesigning the rotor and using a pure helium with higher backing pressures. Overall, our design improves on a previous simulation by Doak et. al. [20] by several orders of magnitude. Another advantage of the focusing element is that by reducing the beam spot size we should be able to increase the beam density by a factor of

30. That would allow for more efficient trapping of atoms or light molecules (provided that we can slow them to sufficiently low velocities) in an optical dipole trap. Note that due to a very short valve opening time as compared to the slow beam propagation time, we expect to have a very high correlation between atom arrival time and velocity. For a beam at 50 m/s allowed to propagate 1 m from the rotor and assuming a timing resolution of 1 ms, we expect better than 2 $\mu$eV energy resolution. Since our beam is not perfectly monochromatic, this would allow us to obtain a high resolution atom-surface interaction spectrum with a single "shot".

We expect to show in the near future slowing of beams composed of light molecules such as $H_2$, $D_2$ and $CH_4$. We shall investigate both atomic and molecular interactions with crystal surfaces at previously unavailable energy ranges with higher resolutions than current techniques allow. As another future application of our slowing method, we would like to mention the possibility of sufficiently slowing helium atoms or light molecules such that they can be trapped in a optical dipole trap for precision spectroscopy.

**Acknowledgments**

We acknowledge support from R.A. Welch Foundation, The Army Research office and the National Science Foundation. We acknowledge assistance from A. Khajetoorians, The Center for Nano and Molecular Science and Technology-University of Texas at Austin, The Center for Electromechanics-University of Texas at Austin and the University of Texas at Austin physics department machine shop. We thank the following people for discussions: D. MacLaren, W. Henderson, M. Fink, G.O. Sitz, H. Langhoff, C.K. Shih, and R.B. Doak.

---